\documentclass[12pt]{article}

\usepackage{authblk}
\usepackage{graphicx}% Include figure files
\usepackage{subcaption} %subfigures
\usepackage{epstopdf} % for windows TexStudio

\def\eq#1{Eq.~(\ref{#1})} % Write a equation as Eq. (foo)
\def\eqs#1{Eqs.~(\ref{#1})} % for Eqs. (foo)
\def\eqn#1{(\ref{#1})} % for (foo) only

\graphicspath{{./Figs/}}

\begin{document}
	
	\bibliographystyle{unsrt}

\title{Macroscopic Lattice Boltzmann Method (MacLAB)}

%\centering{\small{\it (08 May 2018)}}}
%\end{center} }

\author{Jian Guo Zhou}

\affil{Department of Computing and Mathematics\\
	Manchester Metropolitan University\\
	Manchester, M1 5GD, UK\\
	J.Zhou@mmu.ac.uk}

\date{}

\maketitle


\begin{abstract}
The birth of the lattice Boltzmann method (LBM) fulfils a dream that simple arithmetic calculations can simulate complex fluid flows without solving complicated partial differential flow equations. Its power and potential of resolving more and more challenging physical problems have been and are being demonstrated in science and engineering covering a wide range of disciplines such as physics, chemistry, biology, material science and image analysis. The method is a highly simplified model for fluid flows using a few limited fictitious particles that move one grid at a constant time interval and collide each other at a grid point on uniform lattices, which are the two routine steps for implementation of the method to simulate fluid flows. The former represents fluids movement and the latter provides fluid viscosity for diffusion effect. As such, a real complex particle dynamics is approximated as a regular particle model using three parameters of lattice size, particle speed and collision operator.  A fundamental question is ``Are the two steps  integral to the method or can the three parameters be reduced to one for a minimal lattice Boltzmann method?". Here, I show that the collision step can be removed and the standard LBM can be reformulated into a simple macroscopic lattice Boltzmann method (MacLAB). This model relies on macroscopic physical variables only and is completely defined by one basic parameter of lattice size $\delta x$, bringing the LBM into a precise ``Lattice" Boltzmann method. The viscous effect on flows is naturally embedded through the particle speed, making it an ideal automatic simulator for fluid flows. Three additional advantages compared to the existing LBMs are that (i) physical variables can directly be retained as the boundary conditions; (ii) computational memory are much less required and (iii) the model is unconditional stable. The findings have been demonstrated and confirmed with numerical tests including flows that are independent of and dependent on fluid viscosity, 2D and 3D cavity flows, and an unsteady Taylor-Green vortex flow.  This provides an efficient and  powerful model for resolving physical problems in various disciplines of science and engineering.	
\end{abstract}

The LBM is characterised by its simplicity, parallel processing, and easy treatment of boundary conditions \cite{ChenDoolen:1998}.  The first fully discrete model for fluid flows on a square lattice was proposed by 
Hardy et al. \cite{Hardy.etc:1976} in 1976. Ten years later, 
Frisch et al. \cite{Frisch.etc:1986} for the first time obtained a correct lattice gas automata 
(LGA) for Navier-Stokes equations using six-velocity hexagonal lattice. 
The LGA comprises two steps: streaming and collision. The two steps are represented by lattice size and a collision operator on a uniform lattice. In physics, the former and the latter simulate the phenomena of fluid movement and diffusion, respectively, which determine the basic feature of a LGA. 
Often, simulations generated using a LGA are very noisy due to
its Boolean variable with one for the presence and zero for the absence of particles \cite{Chopard.etc:1998,Rivet.etc:2001}. Also, the numerical procedure
involves calculations of particle probability, which reduces the efficiency of the model. To overcome these, the lattice Boltzmann method was proposed \cite{McNamara.etc:1988} and its basic difference
from the LGA is that the Boolean variable is replaced with a particle distribution function. Such approach eliminates the statistical noise in a LGA and retains all the advantages of locality in the kinetic form of a LGA \cite{ChenDoolen:1998}.  McNamara and Zanetti \cite{McNamara.etc:1988} first used the lattice Boltzmann
method as an alternative to the LGA in 1988. As the collision operator takes a complex matrix form, this prevents the LBM from becoming a competing computational method. A breakthrough progress has been made by Higuera and Jim\'{e}nez \cite{Higuera.etc:1989} who linearized the collision term around its local equilibrium state. This greatly simplifies the collision operator. Noble et al. \cite{Noble.etc:1995} used this idea to express the collision operator as  $\Omega_{\alpha \beta} (f_\beta^{eq}-f_\beta)$, in which $f_\beta$ is the particle distribution function; $f_\beta^{eq}$ is the local equilibrium distribution function; and $\Omega_{\alpha \beta}$ is a collision matrix. 
Later, several researchers \cite{Qian:1990,Chen.etc:1991} suggested a simple linearized
form for the collision matrix by using a single time
relaxation towards the local equilibrium distribution, $\Omega_{\alpha \beta}
=- \delta_{\alpha \beta} / \tau$, which is the
Bhatnagar-Gross-Krook \cite{Bhatnagar.etc:1954} collision operator.  In the operator, $\delta_{\alpha \beta} $ is the Kronecker delta function taking one when $\alpha = \beta $, or otherwise zero, and $\tau$ is called the single relaxation time taking a constant and is related to fluid viscosity. This leads to the single relaxation time lattice Boltzman method (SRT LBM) that has been most efficient and used so far, 
\begin{equation}
f_\alpha(x_j + e_{\alpha j} \delta t, t + \delta t) 
= f_\alpha(x_j, t)  + 
\frac{1}{\tau}  [f_\alpha^{eq}(x_j, t)-f_\alpha(x_j , t)],
\label{lb.1} \end{equation}
where $x_j$ is a lattice coordinate along $j$-axis in Cartesian coordinate system, e.g., $j=x,\ y$ in the two dimensional space; $t$ is time; $e_{\alpha j}$ is the $j^{th}$ component of the particle velocity vector ${\bf e}_\alpha$ in $\alpha-$link of the lattice and defined by time step $\delta t$ and lattice size $\delta x$, e.g., ${\bf e}_\alpha = (0,0),\ (e,0),\ (0,e),\  (-e,0),\ (0,-e), \ (e,e),\ (-e,e),\ (-e,-e),\ (e,-e)$ when $\alpha =0 - 8$ for nine particles moving in the two dimensional uniform square lattice (D2Q9), in which $e$ is the particle speed and defined as $e=\delta x/\delta t$; and $f_\alpha^{eq}$ is the local equilibrium distribution function given by
\begin{equation}
f_\alpha^{eq} = 
w_\alpha \rho \left( 1+3 \frac{e_{\alpha i}u_i}{e^2}
+ \frac{9}{2} \frac{e_{\alpha i}e_{\alpha j}u_i u_j}{e^4}
- \frac{3}{2} \frac{u_i u_i}{e^2}  \right),
\label{feq-full}
\end{equation} 
in which $\rho$ is the fluid density and $w_\alpha$ is a weighting factor depending on lattice pattern, e.g., $w_\alpha = 4/9$ when $\alpha = 0$, $w_\alpha = 1/9$ when $\alpha = 1 - 4$ and $w_\alpha = 1/36$ when $\alpha = 5 - 8$ on D2Q9. After the distribution function is calculated from the lattice Botlzmann equation \eqn{lb.1}, the macroscopic phsycal variables, density and velocity are simply updated as
\begin{equation}
\rho (x_j, t) =\sum_\alpha f_\alpha  (x_j, t), \hspace{13mm}
u_i(x_j, t) = \frac{1}{\rho(x_j, t)} \sum_\alpha e_{\alpha i} f_\alpha (x_j, t).
\label{fea-0}
\end{equation}
Since then, the study on the lattice Boltzmann method and applications of the method have received extensive attentions, making it become a very powerful modelling tool in many areas such as thermodynamics \cite{Zarghami:2017}, aerodynamics \cite{Mohamad_Masoud:2014}, multiphase flows \cite{PhysRevE.47.1815}, turbulent flows \cite{Chen.etc:2003}, hemodynamics \cite{Golbert.etc.:2012}, biomechanics \cite{Javed.etc:2017}, image analysis \cite{Chen.etc:2014}, biology \cite{Finck.etc:2014}, environmental science \cite{Zhou.etc:2016}. 

In applications, it is found that the SRT LBM suffers from a numerical instability.  To remedy this, the multiple-relaxation-time (MRT) collision operator was introduced in 1992 \cite{dHumieres:1992, Lallemand.etc:2000}.  This improves the stability but it reduces the efficiency.  To accelerate simulation, a two-relaxation-time collision operator was developed in 2008 \cite{Ginzburg.etc:2008}, which has almost the same efficiency as the SRT LBM. As research progresses, it is noticed that MRT or TRT still suffers from numerical instabilities when fluid flows with very small viscosity are simulated.  After realising that this is caused by an insufficient degree of Galilean invariance in the collision step, Geier et al. \cite{Geier.etc:2006} proposed a cascaded lattice Boltzmann method by relaxaing the particle distribution function to its local equilibrium state in the central moment space, making the LBM stable for simulating flows with small viscosity close to zero.  In 2015, Geier et al. \cite{Geier.etc:2015} further improved their central moment LBM using the cumulant in collision operator, which is called the cumulant lattice Boltzmann method (CLBM). Despite such enhancements, these schemes are more complicated and computational efficiency is reduced as manipulation of matrix is involved. In addition, they share the same drawback as an existing LBM in that the boundary conditions for a physical variable such as velocity cannot be implemented without being converted to particle distribution functions, which further reduces efficiency and accuracy of the methods. Recently, Chen et al. \cite{Chen.etc:2017} developed a simplified lattice Boltzmann method without evolution of particle distribution function, which successfully removes this drawback and enables a direct use of a physical variable as boundary conditions. However, the method involves the two steps of predictor and corrector, which is more complicated than the SRT LBM. Nevertheless, through all these research, the LBM has greatly been improved and developed to a point where it has become a very efficient and flexible alternative numerical method in computational physics. Its potential power is much beyond the original scope, being explored and demonstrated in various disciplines of science and engineering with time \cite{Succi:2001,Wolf:2000,Guo-Shu:2013,Kruger.etc:2017}.

The above literature review highlights that the major research to improve the method have been carried out on the collision operator  for resolving the well-known instability problem in the SRT LBM since 1992, leading to four representative variants of the method, MRT, TRT, Central moment and Cumulant LBMs. Even so, choosing suitable parameters or values for the collision operators is not clear and they cannot be tuned without using trial and error during simulations, which becomes more complicated in a non-single relaxation time scheme
due to its complexity. This unnecessarily wastes time and computing resource, and it may become awkward to simulate a large-scale flow system, preventing the LBM from becoming an automatic simulator for any scale flows when a super-fast computer such as a quantum computer becomes available one day.  In principle, ``Everything should be made as simple as possible, but not simpler" (Albert Einstein's quote). 
Since the central problem comes from the collision operator, the problem may be resolved forever if the collision operator can be removed. Due to the fact that the function of the collision is to relax the distribution function to its local equilibrium state, one may remove the collision operator and use the local equilibrium distribution function to replace the collision by setting $\tau = 1$ in \eq{lb.1}. Following this idea, after mathematical manipulations, I obtain the following simple macroscopic lattice Boltzmann model (MacLAB),
\begin{equation}
\rho (x_j , t)= \sum f_\alpha^{eq}(x_j - e_{\alpha j} \delta t, t-\delta t),
\label{mlb-p.1}
\end{equation}
\begin{equation}
u_i (x_j , t)= \frac{1}{\rho(x_j, t)}  \sum e_{\alpha j}  f_\alpha^{eq}(x_j - e_{\alpha j} \delta t, t-\delta t), 
\label{mlb-u.1}
\end{equation}
to determine physical density and velocity directly from the local equilibrium distribution function without calculating the distribution function using \eq{lb.1} unlike existing LBMs (The theoretical detail can be found in Methods).
Apparently, the model involves the local equilibrium distribution function only. However, doing so brings a new problem of how to consider the fluid viscosity in the absence of collision operator. This can be overcome from the findng through the recovery of the Navier-Stokes equations from \eqs{mlb-p.1} and \eqn{mlb-u.1} that 
if the particle speed $e$ is determined using
\begin{equation}
e = 6 {\nu}/{ \delta x },
\label{mlb-viscosity}
\end{equation}
and is employed in \eq{feq-full} instead of using $e=\delta x/\delta t$ to calculate the local equilirium distribution function $f_\alpha^{eq}$, the flow viscosity is naturally taken into account in the model. In this case, once a lattice size $\delta x$ is chosen, the model is ready to simulate a flow with a viscosity $\nu$ because
$(x_j - e_{\alpha j} \delta t)$ stands for a neighbouring lattice point; $f_\alpha^{eq}$ at time of $(t-\delta t)$ represents its known quantity at the current time; and the particle speed $e$ is determined from \eq{mlb-viscosity}  for use in computation of $f_\alpha^{eq}$. In addition, the time step $\delta t$ is no longer an independent parameter but is calculated as $\delta t = \delta x/e$, which is used in simulations of unsteady flows. 
Consequently, only the lattice  size $\delta x$ is required in the MacLAB for simulation of fluid flows,
bringing the LBM into a precise ``Lattice'' Boltzmann method. This enables the model to become an automatic simulator for any scale flows without tuning other simulation parameters, making it possible and easy to model a large flow system when a super-fast computer such as a quantum computer becomes available in the future. The model is unconditionally stable as it shares the same valid condition as that for $f_\alpha^{eq}$, or the Mack number $M=U_c/e$ is much smaller than 1, in which $U_c$ is a characteristic flow speed. The Mack number can also be expressed as a lattice Reynolds number of $R_{le}=U_c \delta x/ \nu$ via \eq{mlb-viscosity}. In practical simulations, it is found that the model is stable if $R_{le}=U_m \delta x/ \nu < 1$ where $U_m$ is the maximum flow speed and is used as the characteristic flow speed. The main features are that there is no collision operator and only macroscopic physical variables such as density and velocity are required, which are directly used as boundary conditions with a minimum memory requirement. The implementation of the model starting from the initial density and velocity is to (i) choose the lattice size $\delta x$ and determine the particle speed $e$ from \eq{mlb-viscosity}, (ii) calculate $f_\alpha^{eq}$ from \eq{feq-full} using density and velocity, (iii) update the density and velocity using \eqs{mlb-p.1} and \eqn{mlb-u.1}, (iv) apply the boundary conditions, and (v) repeat Step (ii) until a solution is reached. The only limitation of the described model is that, for very small viscosity or high speed flow, the chosen lattice size after satisfying $R_{le} < 1$ may turn out to generate very large lattice points (Lattice points, e.g., for one dimension with length of $L$ is calculated as $N_L=L/\delta x$ and $N_L$ is the lattice points); if the total lattice points is too big such that the demanding computations is beyond the current power of a computer, the simulation cannot be carried out.  Such difficulties may be solved or relaxed through parallel computing using computer techniques such as GPU processors and multiple servers, and will largely or completely removed using quantum computing when a quantum computer becomes available.    

\begin{figure}[h]
	\begin{center}
		\includegraphics[width=2.5in]{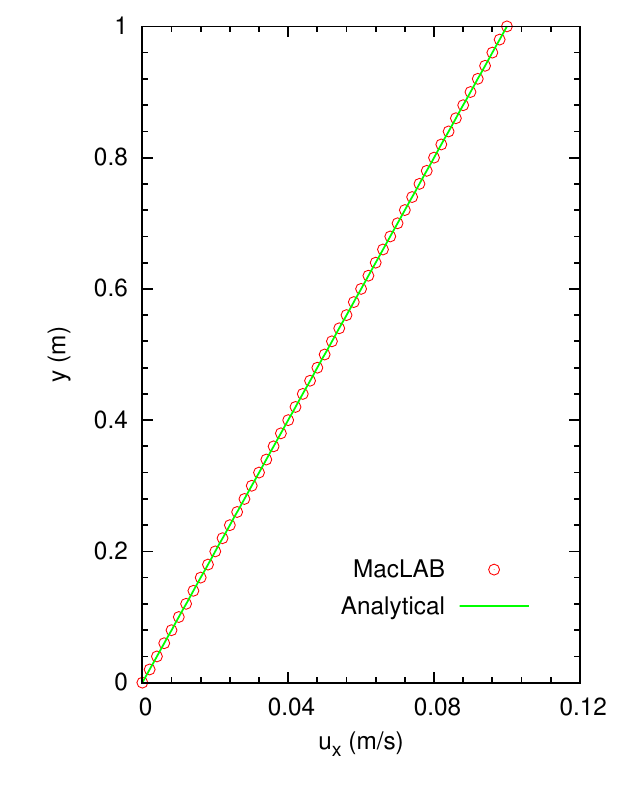}
		\caption{Couette flow through two parallel plates without a pressure gradient. The distance between the plates is $h=1$. The top plate moves at velocity of $0.1$ in the streamwise dierection and the bottom plate is fixed where no calculations are required. The period boundary conditions are applied at the inflow and outflow boundaries. $\delta x = 0.02$ is used for three simulations of flows with three kinematic viscosities of $\nu_1 =0.01,\ \nu_2=0.001$ and $\nu_3=0.0006$, respectively. All the steady numerical results are almost identical and are independent of flow viscosity as shown here in the comparison of one numerical results with the analytical solution.}
		\label{Couette_flow_noPress}
	\end{center}
\end{figure}

In order to demonstrate the validation of the described model, I have carried out five numerical simulations using D2Q9 and D3Q19 lattices for 2D and 3D flows, respectively. For D3Q19, the particle velocity vector is ${\bf e}_\alpha = (e_{\alpha x},e_{\alpha y},e_{\alpha z}) = (0,0,0),\ (e,0,0),\ (-e,0,0), \ (0,e,0), \ (0,-e,0), \linebreak \ (0,0,e),  \ (0,0,-e), \ (e,e,0),\ (-e,-e,0), \ (-e,e,0),\ (e,-e,0),\ (0,e,e), \ (0,-e,-e),  \ (0,-e,e), \linebreak \ (0,e,-e), \ (e,0,e), \ (-e,0,-e), \ (-e,0,e),\ (e,0,-e)$, and the weighting factor $w_\alpha$ for \eq{feq-full}  is  $w_\alpha = 1/3$ when $\alpha = 0$, $w_\alpha = 1/18$ when $\alpha = 1 - 6$ and $w_\alpha = 1/36$ when $\alpha = 7-18$.
The SI units are used with $\rho=1$ in the numerical simulations. 
%\subsection{Couette flow}
The first test is a Couette flow through two parallel plates without a pressure gradient. The distance between the plates is $h=1$. The top plate moves at velocity of $u_x=u_0=0.1$ in the streamwise dierection and the bottom plate is fixed. If $x$ stands for the streamwise direction  and $y$ for the vertical direction, the analytical solution  is
\begin{equation}
u_x(y) = \frac{u_0}{h} y,
\label{Couette -no-pre}
\end{equation}
which is the same test as that used by Chen et al. \cite{Chen.etc:2017}.  This is a very interested case as the steady flow is independent of flow viscosity according to the theory \eqn{Couette -no-pre}. I use $\delta x = 0.02$ and $20 \times 50$ lattices in $x$ and $y$ directions for three simulations of flows with three kinematic viscosities of $\nu_1 =0.01,\ \nu_2=0.001$ and $\nu_3=0.0006$, respectively. The period boundary conditions are applied at the inflow and outflow boundaries.  After the steady solutions are reached, the results are indeed independent of the viscosities and one of those are shown in Fig.~\ref{Couette_flow_noPress}, demonstrating excellent agreement with the analytical solution.

\begin{figure}[h]
	\begin{center}
		\includegraphics[width=2.5in]{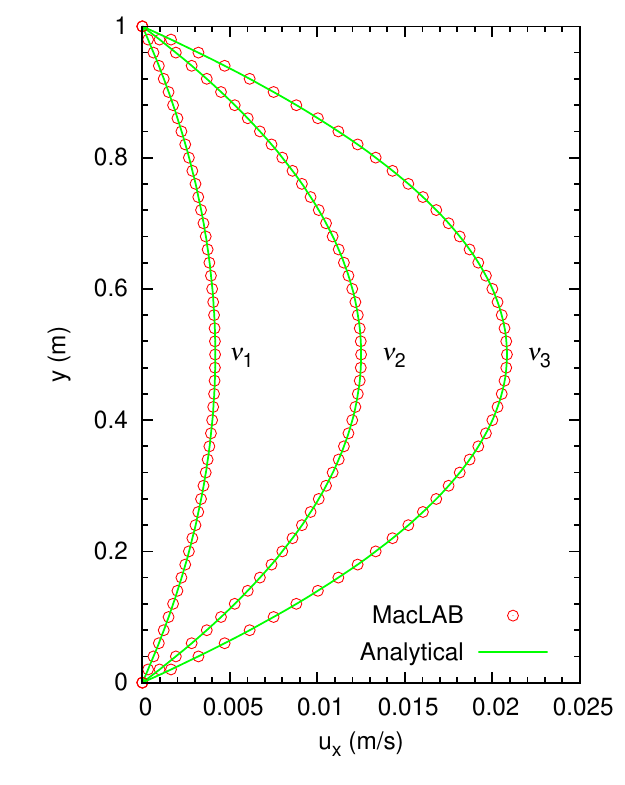}
		\caption{Couette flow through two parallel plates with a pressure gradient of $\partial p/\partial x = -0.0001$. The distance between the plates is $h=1$. Both plates are fixed with zero velocities at top and bottom boundaries where no calculations are needed. The steady numerical results are dependent on flow viscosity as confirmed in the simulations using the three viscosities of $\nu_1 =0.003,\ \nu_2=0.001,$ and $\nu_3 = 0.0006$. }
		\label{Couette_flow_Press_com_nu}
	\end{center}
\end{figure}

The second test is the same flow as that in Test 1 except that a pressure gradient of $\partial p/\partial x = -0.0001$ is specified, which is added to the right hand side of \eq{mlb-u.1} as $+ \delta x/(e \rho) \partial p/\partial x$ \cite{zhou.bk.2004}.  Both plates are fixed with zero velocities at top and bottom boundaries at which no calculations are needed.  The flow is affected by viscosity and the analytical solution is
\begin{equation}
u(y) = \frac{u_0}{h} y + \frac{1}{2\rho \nu} \frac{\partial p}{\partial x} (y^2 - hy).
\label{Couette-PreGrad}
\end{equation}
I simulate this flow using three viscosities of $\nu_1 =0.003,\ \nu_2=0.001,$ and $\nu_3 = 0.0006$.  The numerical results have been plotted in Fig.~\ref{Couette_flow_Press_com_nu}, showing the effect of viscosity on the flow in excellent agreements with the analytical solutions.  This confirms the unique feature that the model can simulate viscous flow correctly due to the use of \eq{mlb-viscosity} although no explicit effect of viscosity on flows is taken into account.

\begin{figure}[h]
	\begin{center}
		\includegraphics[width=3.5in]{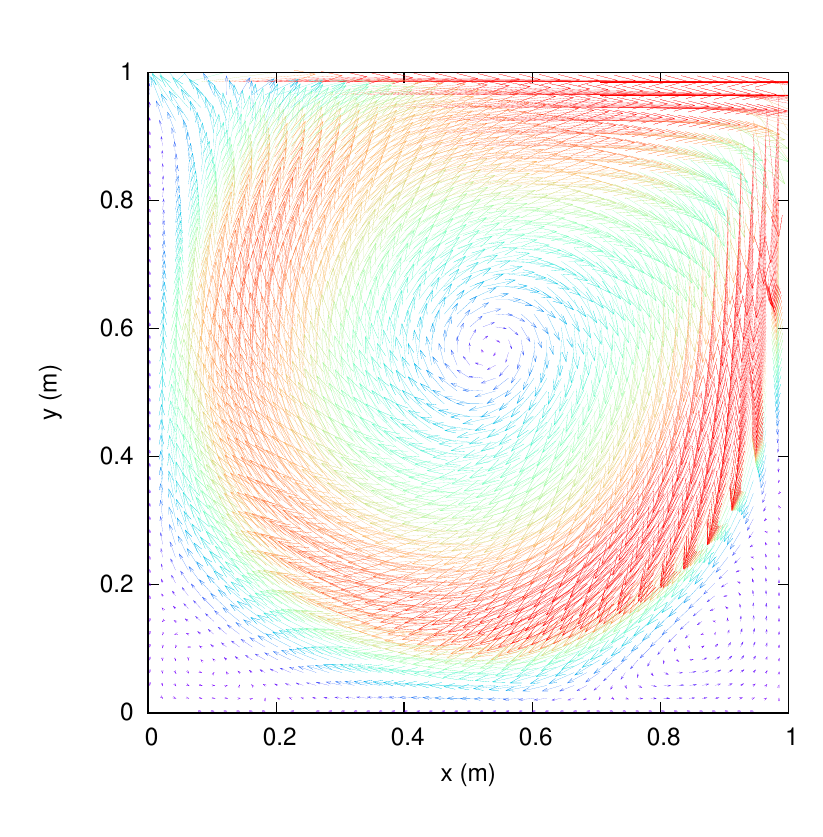}
		\caption{2D cavity flow within $1 \times 1$ square for $R_e = 1000$. The top lid moves at velocity of $u_x = 1$ and  $u_y = 0$ and the other three sides are fixed, or no slip boundary condition is applied. After the steady solution is obtained, the flow pattern in velocity vectors shows a primary vortex and two secondary vortices.}
		\label{2D_cavity_vec}
	\end{center}
\end{figure}

\begin{figure}[h]
	\begin{center}
		\includegraphics[width=3.5in]{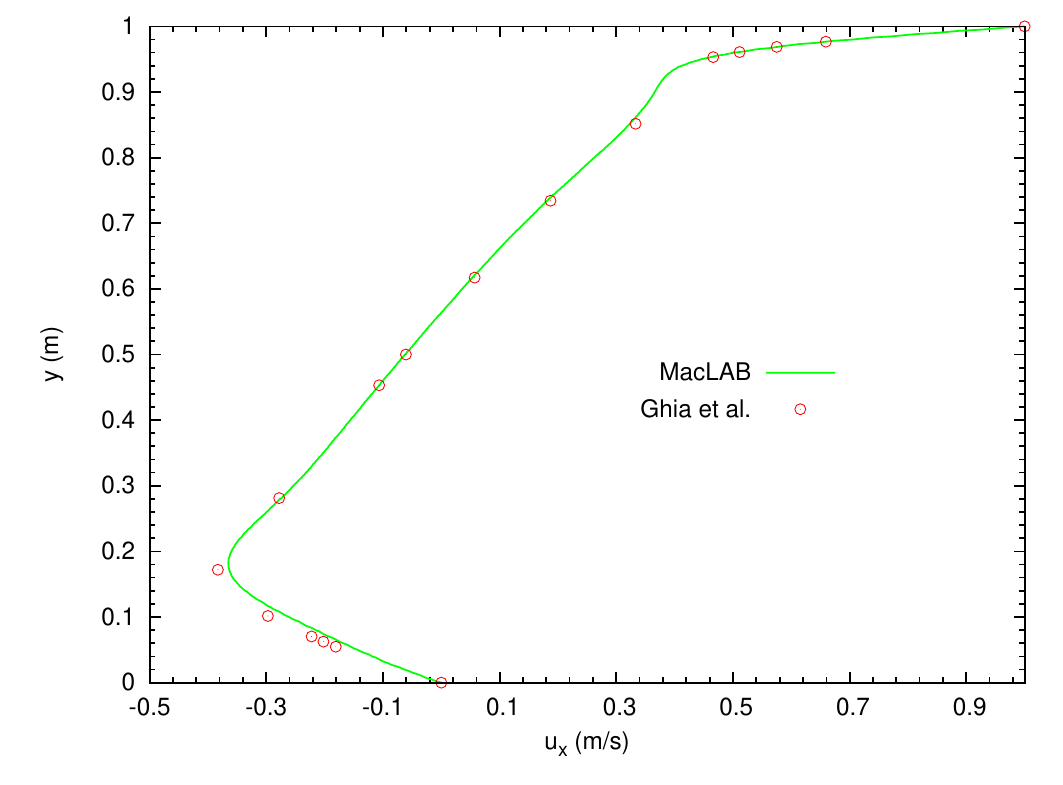}
		\caption{2D cavity flow within $1 \times 1$ square for $R_e = 1000$. The top lid moves at velocity of $u_x = 1$ and  $u_y = 0$ and the other three sides are fixed, or no slip boundary condition is applied. After the steady solution is obtained, the comparison of velocity $u_x$ profile along $y$ direction through the geometric centre of the cavity with the numerical solution by Ghia et al. \cite{Ghia.etc:1982}. }
		\label{2D_cavity_dis-u}
	\end{center}
\end{figure}

\begin{figure}[h]
	\begin{center}
		\includegraphics[width=3.5in]{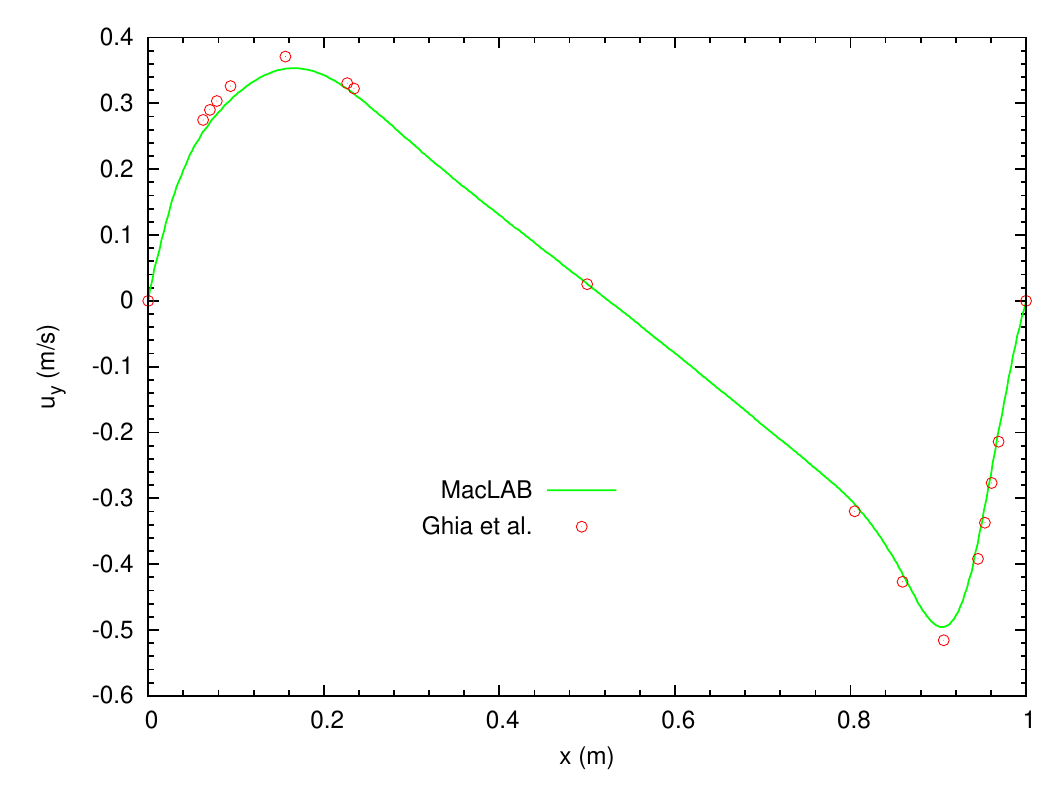}
		\caption{2D cavity flow within $1 \times 1$ square for $R_e = 1000$. The top lid moves at velocity of $u_x = 1$ and  $u_y = 0$ and the other three sides are fixed, or no slip boundary condition is applied. After the steady solution is obtained, the comparison of velocity $u_y$ profile along $x$ direction through the geometric centre of the cavity with the numerical solution by Ghia et al. \cite{Ghia.etc:1982}. }
		\label{2D_cavity_dis-v}
	\end{center}
\end{figure}

The third test is a 2D cavity flow, which is a well-known complex flow within a simple geometry.  The domain is a $1 \times 1$ square.  The boundary conditions are that the top lid moves at velocity of $u_x = u_0$ and  $u_y = 0$ with $u_0=1$; the other three sides are fixed, or no slip boundary condition is applied, i.e., $u_x =0$ and $u_y = 0$. The Reynolds number $R_e = u_0/\nu = 1000$.  I use $\delta x = 0.0025$ or $400 \times 400$ lattices in the simulation, which is carried out on the inside of the cavity excluding the four sides where velocities are retained as boundary conditions. After the steady solution is obtained, the flow pattern in velocity vectors is shown in Fig~\ref{2D_cavity_vec}, which closely agrees with the well-known study by Ghia et al. \cite{Ghia.etc:1982}. The results are further compared against their numerical solution for velocity profiles of $u_x$ and $u_y$ along $y$ and $x$ directions through the geometric centre of the cavity in Figs.~\ref{2D_cavity_dis-u} and \ref{2D_cavity_dis-v}, respectively, demonstraing very good agreements.

\begin{figure}[h]
	\begin{center}
		\includegraphics[width=3.5in]{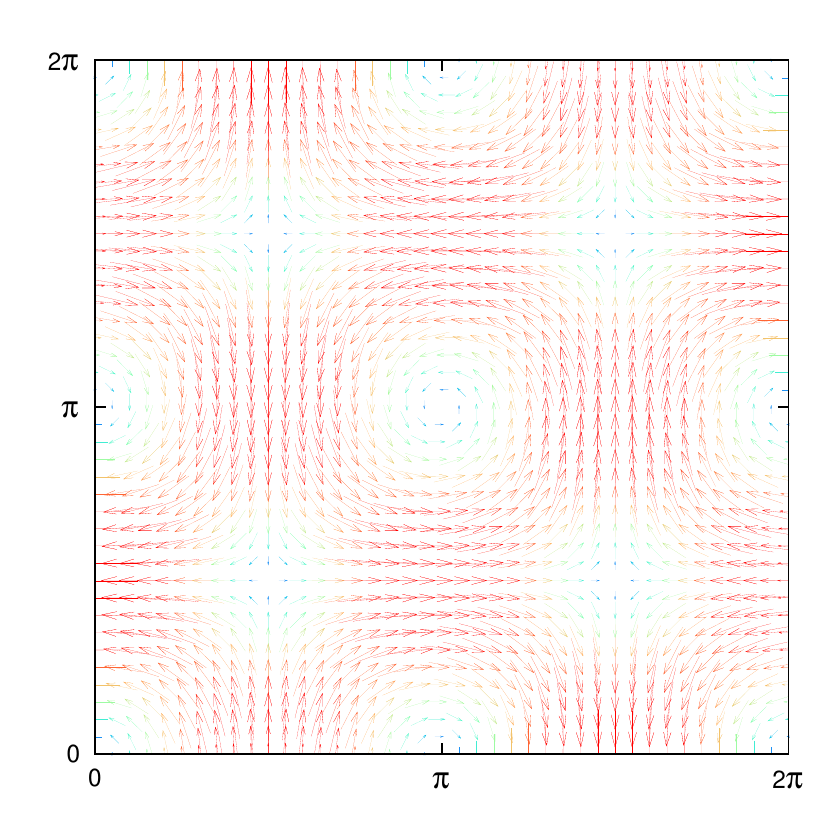}
		\caption{Taylor-Green vortex within $2\pi \times 2\pi$ domain for $R_e = 10$. The initial conditions are $u_x(x,y,0)=-u_0 \cos(x) \sin(y)$ and $u_y(x,y,0)=u_0 \sin(x) \cos(y)$ with $u_0=0.05$. The periodic boundary conditions are used. Here shown is the flow
			pattern in velocity vectors at $t = 30$ seconds, remaining the same vortex pattern as that at initial state.}
		\label{TGV_vec_col.pdf}
	\end{center}
\end{figure}

\begin{figure}[h]
	\begin{center}
		\includegraphics[width=3.6in]{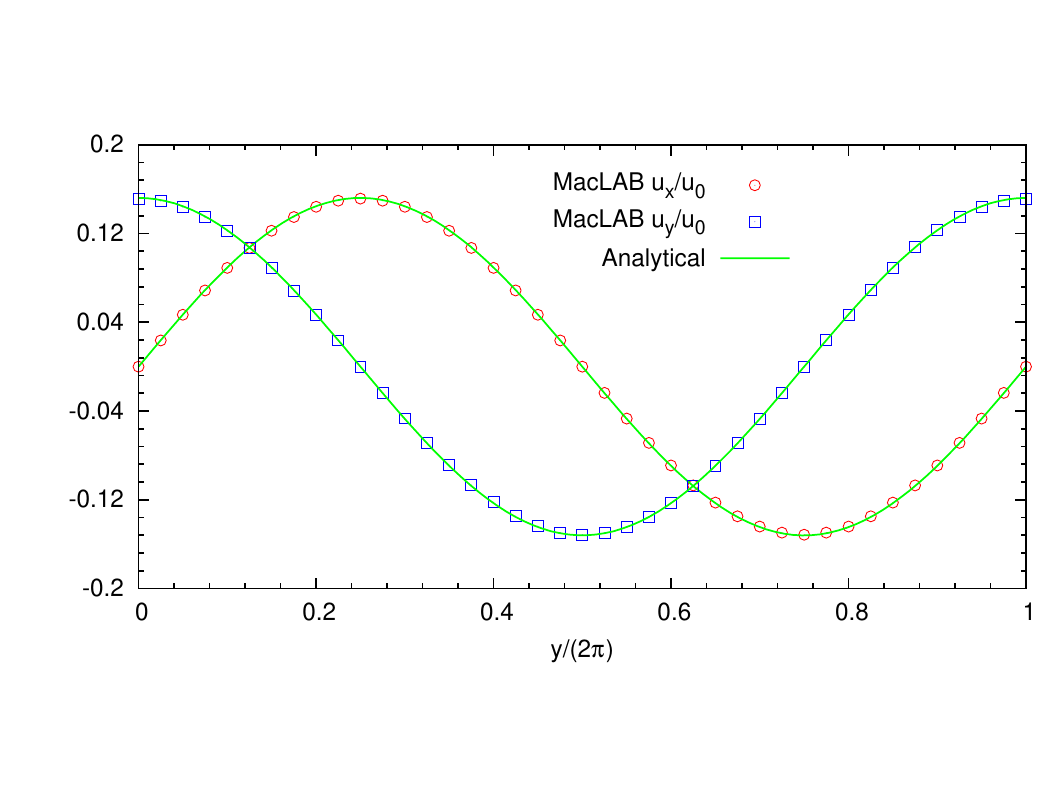}
		\caption{Taylor-Green vortex within $2\pi \times 2\pi$ domain for $R_e = 10$. The initial conditions are $u_x(x,y,0)=-u_0 \cos(x) \sin(y)$ and $u_y(x,y,0)=u_0 \sin(x) \cos(y)$ with $u_0=0.05$. The periodic boundary conditions are used. Here shown is 
			the comparisons of the relative velocity profiles for $u_x/u_0$ at $x=\pi$ and $u_y/u_0$ at $x=\pi/2$ matching the analytical solutions of $u_x(x,y,t)=-u_0 \cos(x) \sin(y) \exp(-2\nu t)$ and $u_y(x,y,t)=u_0 \sin(x) \cos(y) \exp(-2\nu t)$ at $t=30$ seconds.}
		\label{TGV_com_uv_col.pdf}
	\end{center}
\end{figure}

The fourth test is a 2D Taylor-Green vortex. This is an
unsteady flow driven by decaying vortexes for which there is an exact solution of the incompressible Navier-Stokes equations and it is often applied to validation of a numerical method for solution to the incompressible Navier-Stokes equations. 
The initial conditions are $u_x(x,y,0)=-u_0 \cos(x) \sin(y)$ and $u_y(x,y,0)=u_0 \sin(x) \cos(y)$. The analytical solution are $u_x(x,y,t)=-u_0 \cos(x) \sin(y) \exp(-2\nu t)$ and \linebreak  $u_y(x,y,t)=u_0 \sin(x) \cos(y) \exp(-2\nu t)$. The time for an unsteady flow from initial state is accumulated by its increase with time step $\delta t$. I use $\delta x = 0.157$ or $40 \times 40$ lattices for square domain of $2 \pi \times 2\pi$ with kinematic viscosity of $\nu=0.0314$ and $u_0=0.05$, which gives the Reynolds number of $R_e=2\pi u_0/\nu=10$. The periodic boundary conditions are used. The simulation is
run for the total time of 30 seconds. The velocity field is
plotted in Fig.~\ref{TGV_vec_col.pdf}, showing correct flow pattern. The velocity profiles for $u_x$ at $x=\pi$ and $u_y$ at $x=\pi /2$ along $y$-direction are depicted and compared with the analytical solutions in Fig.~\ref{TGV_com_uv_col.pdf}, showing excellent agreemeents and confirming the accuracy of the method for an unsteady flow.

\begin{figure}[h]
	\begin{center}
		\includegraphics[width=3.5in]{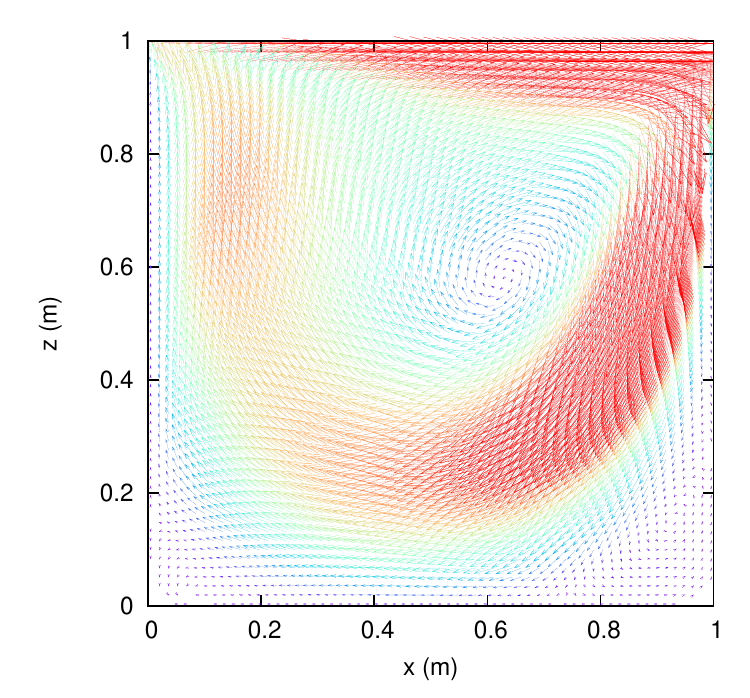}
		\caption{3D cavity flow within $1 \times 1 \times 1$ cube for $R_e = 400$. The top lid moves at velocity of $u_x = 1$,  $u_y = 0$ and $u_z = 0$ and the other five sides are fixed, or no slip boundary condition is applied. After the solution is reached,  the flow pattern in vectors in $x - z$ centroidal plane shows the primary and secondary vortices.}
		\label{3D_Re400_vec_uw_col.pdf}
	\end{center}
\end{figure}

\begin{figure}[h]
	\begin{center}
		\includegraphics[width=3.5in]{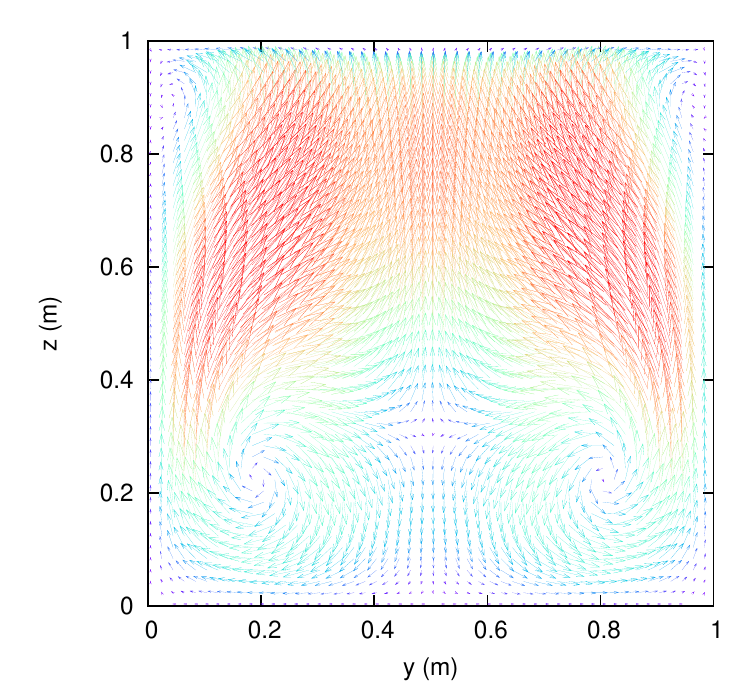}
		\caption{3D cavity flow within $1 \times 1 \times 1$ cube for $R_e = 400$. The top lid moves at velocity of $u_x = 1$,  $u_y = 0$ and $u_z = 0$ and the other five sides are fixed, or no slip boundary condition is applied. After the solution is reached, the flow pattern in vectors in $y - z$ centroidal plane shows one pair of strong secondary vortices at bottom and one pair of weak secondary vortices at top. }
		\label{3D_Re400_vec_vw_col.pdf}
	\end{center}
\end{figure}

\begin{figure}[h]
	\begin{center}
		\includegraphics[width=3.5in]{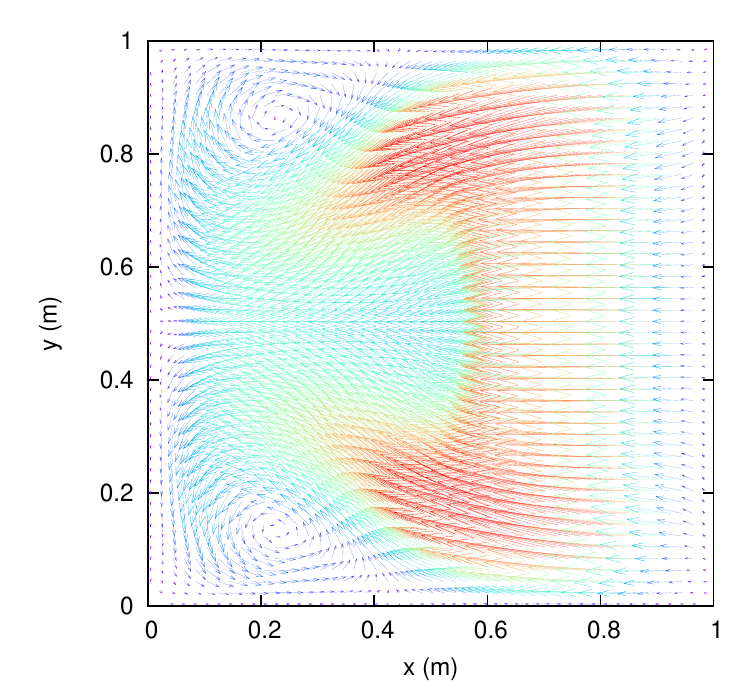}
		\caption{3D cavity flow within $1 \times 1 \times 1$ cube for $R_e = 400$. The top lid moves at velocity of $u_x = 1$,  $u_y = 0$ and $u_z = 0$ and the other five sides are fixed, or no slip boundary condition is applied. After the solution is reached, the flow pattern in vectors in $x - y$ centroidal plane shows a pair of third vortices close to inflow boundary.}
		\label{3D_Re400_vec_uv}
	\end{center}
\end{figure}

\begin{figure}[h]
	\begin{center}
		\includegraphics[width=3.5in]{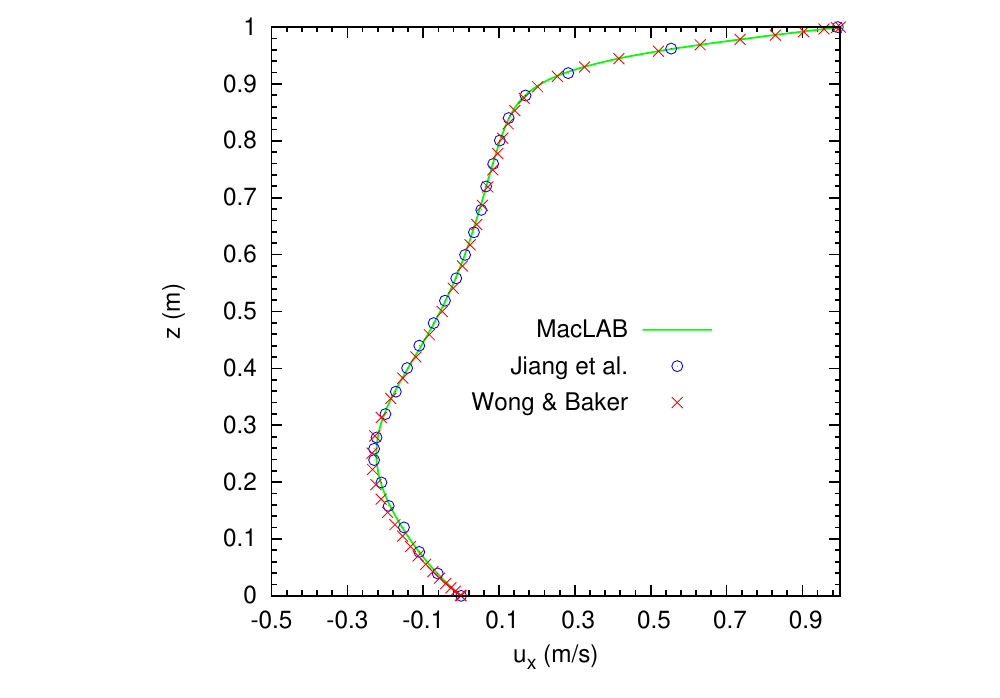}
		\caption{3D cavity flow within $1 \times 1 \times 1$ cube for $R_e = 400$. The top lid moves at velocity of $u_x = 1$,  $u_y = 0$ and $u_z = 0$ and the other five sides are fixed, or no slip boundary condition is applied. After the solution is reached, the comparisons of the distribution of the velocity component $u_x$ on the vertical plane centerline with the restuls by Wong and Baker \cite{Wong.Baker:2002} and Jiang et al. \cite{Jiang.etc:1994}.}
		\label{3D Re400 dis-u-250_col.pdf}
	\end{center}
\end{figure}

The final test is a 3D cavity flow. This is again a well-known complex flow involving 3D vortices within a simple cube with the dimensions of $1 \times 1 \times 1$ in streamwise direction $x$, spanwise direction $y$ and vertical direction $z$.  No-slip boundary conditions, i.e, $u_x=0$ and $u_y=0$ and $u_z=0$, are applied to five fixed sides except for the top lid, where $u_x=u_0$, $u_y=0$ and $u_z=0$ with $u_0=1$ are specified.  The Reynolds number is $R_e = u_0/\nu = 400$.  $\delta x = 0.004$ or total lattices of $250 \times 250 \times 250$  are used and the simulation is undertaken only within the cube excluding the boundaries where the velocities are retained. After the steady solution is reached, the flow characterises are displayed 
through the two dimensional planar projections of the velocity vector field on the $x$-$z$, $y$-$z$ and $x$-$y$ centroidal planes of the cube in Figs.~\ref{3D_Re400_vec_uw_col.pdf}, \ref{3D_Re400_vec_vw_col.pdf} and \ref{3D_Re400_vec_uv}, respectively, demonstrating good agreed flow patterns with those by Wong and Baker \cite{Wong.Baker:2002}. In addition, the distribution of the velocity component $u_x$ on the vertical plane centerline is widely used as a 3D lid-driven cavity benchmark test. I compare this velocity component against the results by Wong and Baker \cite{Wong.Baker:2002} and also by Jiang et al. \cite{Jiang.etc:1994} in Fig.~\ref{3D Re400 dis-u-250_col.pdf}, showning good agreements.

In conclusion, the results demonstrate that the MacLAB is able to simulate fluid flows using only lattice size, bringing the LBM into a precise Lattice Boltzmann method. This takes the research on the method into a new era when future work may focus on improving on accuracy of or formulating a new local equilibrium distribution function. The particle speed is determined through the viscosity and lattice size and the time step $\delta t$ is calculated as $\delta t = \delta x /e$.  The model is unconditional stable as long as the valid condition for the local equilibrium distribution function holds. All these make the method an automatic simulator for fluid flows. The method is straightforward to be extended for resolving other physical problems in different disciplines.

\section*{Figures.}  




\clearpage

\vspace{1cm}
\noindent
{\bf METHODS} 

\noindent
I present the detail of the derivation for the present model.  Setting $\tau = 1$ in \eq{lb.1} leads to
\begin{equation}
f_\alpha(x_j + e_{\alpha j} \delta t, t + \delta t) =f_\alpha^{eq}(x_j, t),
\label{MacLBM.d1} \end{equation}
which can be rewritten as
\begin{equation}
f_\alpha(x_j, t) = f_\alpha^{eq} (x_j - e_{\alpha j} \delta t, t - \delta t).
\label{MacLBM.d2} \end{equation}
Taking $\sum$ \eq{MacLBM.d2} and $\sum e_{\alpha i}$\eq{MacLBM.d2} yields
\begin{equation}
\sum f_\alpha(x_j, t) = \sum f_\alpha^{eq} (x_j - e_{\alpha j} \delta t, t - \delta t),
\label{MacLBM.d3} \end{equation}
and 
\begin{equation}
\sum e_{\alpha i} f_\alpha(x_j, t) = \sum e_{\alpha i} f_\alpha^{eq} (x_j - e_{\alpha j} \delta t, t - \delta t),
\label{MacLBM.d4} \end{equation}
respectively.  In the lattice Boltzmann method, the density and velocity are determined using the distribution function as
\begin{equation}
\rho(x_j, t)=\sum_\alpha f_\alpha (x_j, t), \hspace{13mm}
u_i (x_j, t) = \frac{1}{\rho} \sum_\alpha e_{\alpha i} f_\alpha (x_j, t).
\label{MacLBM.d5} \end{equation}
Combining \eq{MacLBM.d5}  with \eqs{MacLBM.d3} and \eqn{MacLBM.d4} results in the current MacLAB, \eqs{mlb-p.1} and \eqn{mlb-u.1}.   Since the local equilibrium distribution function $f_\alpha^{eq}$ has the features of
\begin{equation}
\sum_\alpha f_\alpha^{eq} (x_j, t) =\rho (x_j, t), \hspace{13mm}
\frac{1}{\rho} \sum_\alpha e_{\alpha i} f_\alpha^{eq} (x_j, t) = u_i (x_j, t),
\label{fea-0}
\end{equation}
with reference to \eq{MacLBM.d5} the following relationships,
\begin{equation}
\sum_\alpha f_\alpha (x_j, t) = \sum_\alpha f_\alpha^{eq} (x_j, t) , \hspace{13mm}
\sum_\alpha e_{\alpha i} f_\alpha (x_j, t) =  \sum_\alpha e_{\alpha i} f_\alpha^{eq} (x_j, t),
\label{fea-0add}
\end{equation}
hold, which are the conditions that retain the conservation of the mass and momentum in the lattice Boltzmann method.

Next, I prove that the continuity and Navier-Stokes equations can be recovered from \eqs{mlb-p.1} and \eqn{mlb-u.1}.  Rewriting \eq{lb.1} as
\begin{equation}
f_\alpha(x_j , t) = f_\alpha(x_j - e_{\alpha j} \delta t, t - \delta t) 
+ \frac{1}{\tau}  [f_\alpha^{eq}(x_j - e_{\alpha j} \delta t, t - \delta t) -f_\alpha(x_j - e_{\alpha j} \delta t, t - \delta t) ].
\label{MacLBM-lb.d1} \end{equation}
Apparently, when $\tau  = 1$, the above equation becomes \eq{MacLBM.d2} that leads to \eqs{mlb-p.1} and \eqn{mlb-u.1}; hence \eq{MacLBM-lb.d1} is a general equation and is used in the following derivation. 
Applying a Taylor expansion to the two terms on the right hand side of \eq{MacLBM-lb.d1} in time and space
at point $({\bf x}, t)$ yields
\begin{equation}
f_\alpha(x_j - e_{\alpha j} \delta t, t - \delta t) 
=
f_\alpha(x_j , t) - \delta t \left ( \frac{\partial}{\partial t}+e_{\alpha j}
\frac{\partial}{\partial x_j} \right ) f_\alpha +
\frac{1}{2} \delta t^2 \left ( \frac{\partial}{\partial t}+e_{\alpha j}
\frac{\partial}{\partial x_j} \right )^2 f_\alpha 
+ {\cal O} (\delta t^3),
\label{lb.3} \end{equation}
and
\begin{equation}
f_\alpha^{eq}(x_j - e_{\alpha j} \delta t, t - \delta t)
=
f_\alpha^{eq}(x_j , t) - \delta t \left ( \frac{\partial}{\partial t}+e_{\alpha j}
\frac{\partial}{\partial x_j} \right ) f_\alpha^{eq} +
\frac{1}{2} \delta t^2 \left ( \frac{\partial}{\partial t}+e_{\alpha j}
\frac{\partial}{\partial x_j} \right )^2 f_\alpha^{eq} 
+ {\cal O} (\delta t^3).
\label{lb.3.1} \end{equation}
According to the Chapman-Enskog analysis, $f_\alpha$ can be expanded around $f_\alpha^{(0)}$ 
\begin{equation}
f_\alpha = f_\alpha^{(0)} +  f_\alpha^{(1)} \delta t +
f_\alpha^{(2)} \delta t^2 + {\cal O} (\delta t^3).
\label{fa-ex.1} \end{equation}
After substituting \eqs{lb.3}, \eqn{lb.3.1} and \eqn{fa-ex.1} into \eq{MacLBM-lb.d1}, equating the coefficients results in for the order $(\delta t)^0$
\begin{equation}
f_\alpha^{(0)} = f_\alpha^{eq},
\label{Chpman-Enskog.01} \end{equation}
for the order $(\delta t)^1$
\begin{equation}
\left ( \frac{\partial}{\partial t}+e_{\alpha j}
\frac{\partial}{\partial x_j} \right ) f_\alpha^{(0)},
=-\frac{1}{\tau} f_\alpha^{(1)} ,
\label{Chpman-Enskog.1} \end{equation}
and for the order $(\delta t)^2$
\begin{equation}
\left ( \frac{\partial}{\partial t}+e_{\alpha j}
\frac{\partial}{\partial x_j} \right ) f_\alpha^{(1)} 
- \frac{1}{2} \left ( \frac{\partial}{\partial t}+e_{\alpha j}
\frac{\partial}{\partial x_j} \right )^2 f_\alpha^{(0)}
=-\frac{1}{\tau} f_\alpha^{(2)} + \frac{1}{\tau} \left ( \frac{\partial}{\partial t}+e_{\alpha j}
\frac{\partial}{\partial x_j} \right ) f_\alpha^{(1)} .
\label{Chpman-Enskog.2} \end{equation}
Substitution of \eq{Chpman-Enskog.1} into the 
above equation gives
\begin{equation}
\left (\frac{\partial}{\partial t}+e_{\alpha j}
\frac{\partial}{\partial x_j} \right ) f_\alpha^{(1)} 
- \frac{1}{2} \left ( \frac{\partial}{\partial t}+e_{\alpha j}
\frac{\partial}{\partial x_j} \right ) 
(-\frac{1}{\tau} f_\alpha^{(1)})
=-\frac{1}{\tau} f_\alpha^{(2)} + \frac{1}{\tau} \left ( \frac{\partial}{\partial t}+e_{\alpha j}
\frac{\partial}{\partial x_j} \right ) f_\alpha^{(1)} ,
\label{Chpman-Enskog.3} \end{equation}
which is rearranged as
\begin{equation}
\left ( 1- \frac{1}{2 \tau} \right ) \left (\frac{\partial}{\partial t}+e_{\alpha j}
\frac{\partial}{\partial x_j} \right) 
f_\alpha^{(1)}
=-\frac{1}{\tau} f_\alpha^{(2)} .
\label{Chpman-Enskog.4} \end{equation}
From \eq{Chpman-Enskog.1} +  
\eq{Chpman-Enskog.4} $ \times \delta t$, I have
\begin{equation}
\left (\frac{\partial}{\partial t}+e_{\alpha j}
\frac{\partial}{\partial x_j} \right )  f_\alpha^{(0)} 
+ \delta t \left ( 1- \frac{1}{2 \tau} \right ) \left ( \frac{\partial}{\partial t}+e_{\alpha j}
\frac{\partial}{\partial x_j} \right ) 
f_\alpha^{(1)}
= 
-\frac{1}{\tau} (f_\alpha^{(1)}+\delta t f_\alpha^{(2)}) .
\label{Chpman-Enskog.5} \end{equation}
Now taking $\sum$\eq{Chpman-Enskog.5} leads to
\begin{equation}
\sum  \left (\frac{\partial}{\partial t}+e_{\alpha j}
\frac{\partial}{\partial x_j} \right ) f_\alpha^{(0)} 
= 0
\label{Chpman-Enskog.6} \end{equation}
as
\begin{equation}
\sum_\alpha f_\alpha^{(1)} = \sum_\alpha f_\alpha^{(2)} = \sum_\alpha e_{\alpha i} f_\alpha^{(1)} = 
\sum_\alpha e_{\alpha i} f_\alpha^{(2)} =  0
\label{relation-chapman}
\end{equation}
due to the condition of conservation of mass and momentum \eq{fea-0}.  Evaluating the terms in the above equation using \eq{feq-full} produces the exact continuity equation,
\begin{equation}
\frac{\partial \rho}{\partial t} + \frac{\partial (\rho u_j)}{\partial x_j} = 0.
\label{relation-chapman-mass}
\end{equation}
Multipling \eq{Chpman-Enskog.5} by $e_{\alpha i}$ provides
\begin{equation}
\left (\frac{\partial}{\partial t}+e_{\alpha j}
\frac{\partial}{\partial x_j} \right )  e_{\alpha i} f_\alpha^{(0)} 
+ \delta t \left ( 1- \frac{1}{2 \tau} \right ) \left ( \frac{\partial}{\partial t}+e_{\alpha j}
\frac{\partial}{\partial x_j} \right ) 
e_{\alpha i}  f_\alpha^{(1)}
= 
-\frac{1}{\tau} (e_{\alpha i}  f_\alpha^{(1)}+\delta t e_{\alpha i} f_\alpha^{(2)}) .
\label{Chpman-Enskog.7} \end{equation}
Taking $\sum$\eq{Chpman-Enskog.7} leads to
\begin{equation}
\sum  \left (\frac{\partial}{\partial t}+e_{\alpha j}
\frac{\partial}{\partial x_j} \right ) e_{\alpha i} f_\alpha^{(0)} 
+ \delta t \left ( 1- \frac{1}{2 \tau} \right ) \sum  \left ( \frac{\partial}{\partial t}+e_{\alpha j}
\frac{\partial}{\partial x_j} \right ) 
e_{\alpha i}  f_\alpha^{(1)}
= 0 
\label{C-E.Moment.1} \end{equation}
under the same condition \eqn{relation-chapman} as that in the derivation of \eq{Chpman-Enskog.6}.  Evaluating the terms in the above equation using \eq{feq-full} produces the exact momentum equation, the Navier-Stokes equation at second-order accuracy on condition that the Mack number $M=U_c/e << 1$,
\begin{equation}
\frac{\partial (\rho u_i)}{\partial t} + \frac{\partial (\rho u_i u_j)}{\partial x_j} = -\frac{\partial p}{\partial x_i} 
+ \nu \frac{\partial ^2 (\rho u_i)}{\partial x_j^2} ,
\label{relation-chapman-moment}
\end{equation}
where pressure $p$ is defined as
\begin{equation}
p=\frac{1}{3} \rho e^2
\label{relation-chapman-pressure}
\end{equation}
and the kinmatic viscosity is
\begin{equation}
\nu =\frac{1}{6} ( 2 \tau - 1 ) e \delta x.
\label{relation-chapman-viscosity}
\end{equation}
As $\tau$ takes a constant, use of $\tau = 1$ will recovers the continuity and the Navier-Stoker equations at the second-order accurate as the above derivation shows.  In this case,  \eq{relation-chapman-viscosity} becomes \eq{mlb-viscosity}, which determines the particle speed $e$.

\vspace{1cm}
\noindent
{\bf ADDITIONAL INFORMATION}

\begin{figure}[h]
	\begin{subfigure}{.5\textwidth}
		\centering
		\includegraphics[width=2in]{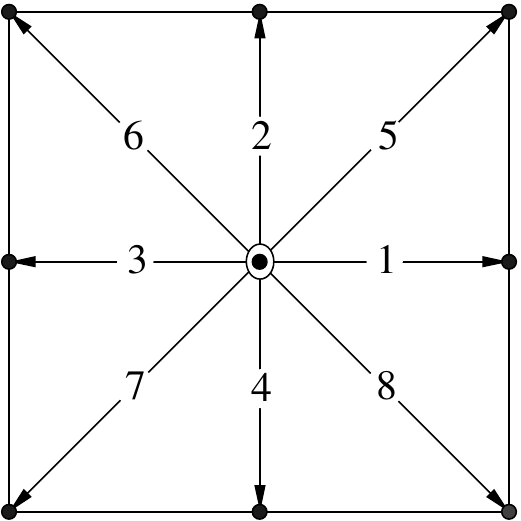}
		\caption{D2Q9 square lattice.}
		\label{SUPPL-D2Q9}
	\end{subfigure}%
	\begin{subfigure}{.5\textwidth}
		\centering
		\includegraphics[width=2.2in]{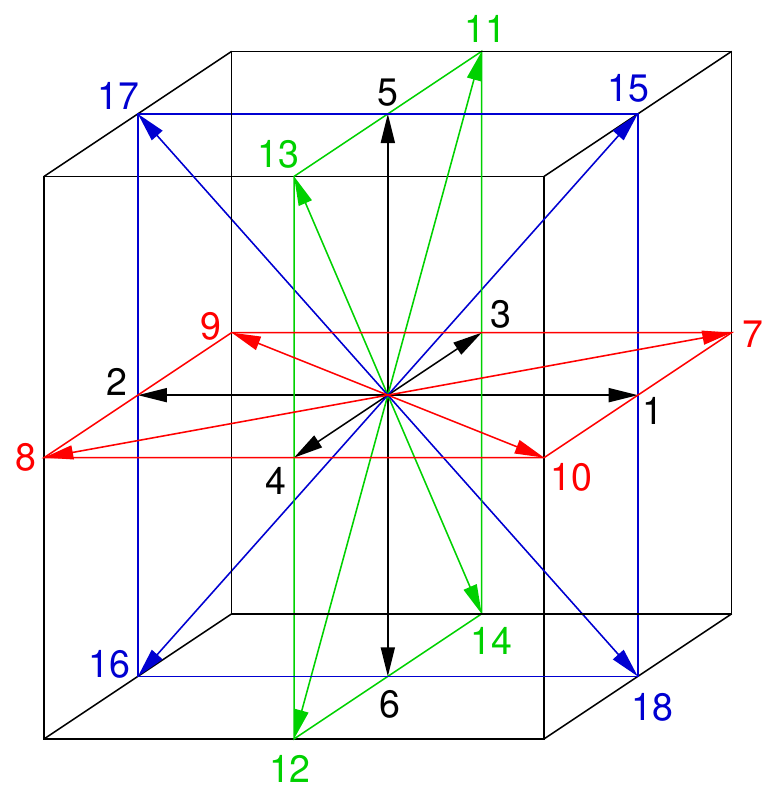}
		\caption{D3Q19 cubic lattice.}
		\label{SUPPL-D3Q19}
	\end{subfigure}
	\caption{Square and cubic Lattices for 2D and 3D flows.}
	\label{D2Q9_D3Q19}
\end{figure}

\noindent
{\bf D2Q9 and D3Q19 Lattice Structures}.
The D2Q9 uniform square and D3Q19 cubic lattices are depicted in Figs.~\ref{SUPPL-D2Q9} and  \ref{SUPPL-D3Q19}, respectively.  

\bibliography{MacLAB}

\begin{thebibliography}{10}

\bibitem{ChenDoolen:1998}
S.~Chen and G.~D. Doolen.
\newblock Lattice {Boltzmann} method for fluid flows.
\newblock {\em Annual Review of Fluid Mechanics}, 30:329--364, 1998.

\bibitem{Hardy.etc:1976}
J.~Hardy, O.~{de Pazzis}, and Y.~Pomeau.
\newblock Molecular dynamics of a classical lattice gas: Transport properties
  and time correlation functions.
\newblock {\em Phys. Rev. A}, 13:1949--1961, 1976.

\bibitem{Frisch.etc:1986}
U.~Frisch, B.~Hasslacher, and Y.~Pomeau.
\newblock Lattice-gas automata for the {Navier-Stokes} equation.
\newblock {\em Physical Review Letters}, 56:1505--1508, 1986.

\bibitem{Chopard.etc:1998}
B.~Chopard and M.~Droz.
\newblock {\em Cellular Automata Modeling of Physical Systems}.
\newblock Cambridge University Press, UK, 1998.

\bibitem{Rivet.etc:2001}
J.~P. Rivet and J.~P. Boon.
\newblock {\em Lattice Gas Hydrodynamics}.
\newblock Cambridge University Press, UK, 2001.

\bibitem{McNamara.etc:1988}
G.~R. McNamara and G.~Zanetti.
\newblock Use of the {Boltzmann} equation to simulate lattice-gas automata.
\newblock {\em Phys. Rev. Lett.}, 61:2332--2335, 1988.

\bibitem{Higuera.etc:1989}
F.~Higuera and J.~Jim\'{e}nez.
\newblock Boltzmann approach to lattice gas simulations.
\newblock {\em Europhys lett.}, 9:663--668, 1989.

\bibitem{Noble.etc:1995}
D.~R. Noble, S.~Chen, J.~G. Georgiadis, and R.~O. Buckius.
\newblock A consistent hydrodynamic boundary condition for the lattice
  {Boltzmann} method.
\newblock {\em Physics of Fluids}, 7:203--209, 1995.

\bibitem{Qian:1990}
Y.~H. Qian.
\newblock {\em Lattice Gas and lattice kinetic theory applied to the
  {Navier-Stokes} equations}.
\newblock PhD thesis, Universit\'{e} Pierre et Marie Curie, Paris, 1990.

\bibitem{Chen.etc:1991}
S.~Chen, H.~D. Chen, D.~Martinez, and W.~Matthaeus.
\newblock Lattice {Boltzmann} model for simulation of magnetohydrodynamics.
\newblock {\em Phys. Rev. Lett.}, 67:3776--3779, 1991.

\bibitem{Bhatnagar.etc:1954}
P.~L. Bhatnagar, E.~P. Gross, and M.~Krook.
\newblock A model for collision processes in gases. i: small amplitude
  processes in charged and neutral one-component system.
\newblock {\em Phys. Rev.}, 94:511--525, 1954.

\bibitem{Zarghami:2017}
Ahad Zarghami and Harry E. A.~Van den Akker.
\newblock Thermohydrodynamics of an evaporating droplet studied using a
  multiphase lattice {Boltzmann} method.
\newblock {\em Physical Review E}, 95:043310, 2017.

\bibitem{Mohamad_Masoud:2014}
H.~H. Mohamad and M.~Masoud.
\newblock Continuous and discrete adjoint approach based on lattice {Boltzmann}
  method in aerodynamic optimization part i: Mathematical derivation of adjoint
  lattice {Boltzmann} equations.
\newblock {\em Advances in Applied Mathematics and Mechanics}, 6(5):570--589,
  2014.

\bibitem{PhysRevE.47.1815}
Xiaowen Shan and Hudong Chen.
\newblock Lattice boltzmann model for simulating flows with multiple phases and
  components.
\newblock {\em Phys. Rev. E}, 47:1815--1819, Mar 1993.

\bibitem{Chen.etc:2003}
Hudong Chen, Satheesh Kandasamy, Steven Orszag, Rick Shock, Sauro Succi, and
  Victor Yakhot.
\newblock Extended {Boltzmann} kinetic equation for turbulent flows.
\newblock {\em Science}, 301:633--636, 2003.

\bibitem{Golbert.etc.:2012}
D.~R. Golbert, P.~J. Blanco, A.~Clausse, and R.~A. Feij\'{o}o.
\newblock Tuning a lattice-{Boltzmann} model for applications in computational
  hemodynamics.
\newblock {\em Medical Engineering \& Physics}, 34:339--349, 2012.

\bibitem{Javed.etc:2017}
Sana Javed, Ayesha Sohail, Khadija Maqbool, Saad~Ihsan Butt, and Qasim~Ali
  Chaudhry.
\newblock The lattice {Boltzmann} method and computational analysis of bone
  dynamics-i.
\newblock {\em Complex Adaptove Systems Modeling}, 5:1--14, 2017.

\bibitem{Chen.etc:2014}
Junhui Chen, Zhenhua Chai, Baochang Shi, and Wenhuan Zhang.
\newblock Lattice {Boltzmann} method for filtering and contour detection of the
  natural images.
\newblock {\em Computers and Mathematics with Applications}, 68:257--268, 2014.

\bibitem{Finck.etc:2014}
M.~Finck, D.~H\"{a}nel, and I.~Wlokas.
\newblock Simulation of nasal flowby lattice {Boltzmann} methods.
\newblock {\em Computers in Biology and Medicine}, 37:739--749, 2007.

\bibitem{Zhou.etc:2016}
J.~G. Zhou, P.~M. Haygarth, P.~J.~A. Withers, C.~J.~A. Macleod, P.~D. Falloon,
  K.~J. Beven, M.~C. Ockenden, K.~J. Forber, M.~J. Hollaway, R.~Evans, A.~L.
  Collins, K.~M. Hiscock, C.~Wearing, R.~Kahana, and M.~L.~Villamizar Velez.
\newblock Lattice {Boltzmann} method for the fractional advection-diffusion
  equation.
\newblock {\em Physical Review E}, 93:043310, 2016.

\bibitem{dHumieres:1992}
D.~d'Humi\`{e}res.
\newblock Generalized lattice {Boltzmann} equations. in rarefied gas dynamics.
\newblock In B.~D. Shizgal and D.~P.Weaver, editors, {\em Rarefied Gas
  Dynamics: Theory and Simulations, Progress in Astronautics and Aeronautics},
  volume 159, pages 450--458. 1992.

\bibitem{Lallemand.etc:2000}
Pierre Lallemand and Li-Shi Luo.
\newblock Theory of the lattice {Boltzmann} method: Dispersion, dissipation,
  isotropy, galilean invariance, and stability.
\newblock {\em Physical Review E}, 61:6546--6562, 2000.

\bibitem{Ginzburg.etc:2008}
Irina Ginzburg, Frederik Verhaeghe, and Dominique d'Humi\`{e}res3.
\newblock Two-relaxation-time lattice {Boltzmann} scheme: About
  parametrization,velocity, pressure andmixed boundary conditions.
\newblock {\em Communications in Computational Physics}, 3(2):427--478, 2008.

\bibitem{Geier.etc:2006}
Martin Geier, Andreas Greiner, and Jan~G. Korvink.
\newblock Cascaded digital lattice {Boltzmann} automata for high reynolds
  number flow.
\newblock {\em Physical Review E}, 73:066705, 2006.

\bibitem{Geier.etc:2015}
Martin Geier, Martin Sch\"{o}nherr, Andrea Pasquali, and Manfred Krafczyk.
\newblock The cumulant lattice {Boltzmann} equation in three dimensions: Theory
  and validation.
\newblock {\em Computers and Mathematics with Applications}, 70:507--547, 2015.

\bibitem{Chen.etc:2017}
Z.~Chen, C.~Shu, Y.~Wang, L.~M. Yang, and D.~Tan.
\newblock A simplified lattice {Boltzmann} method without evolution of
  distribution function.
\newblock {\em Advances in Applied Mathematics and Mechanics}, 9:1--22, 2017.

\bibitem{Succi:2001}
Sauro Succi.
\newblock {\em The Lattice Boltzmann Equation for Fluid Dynamics and Beyond}.
\newblock Oxford University Press, 2001.

\bibitem{Wolf:2000}
Dieter Wolf-Gladrow.
\newblock {\em Lattice-Gas Cellular Automata and Lattice Boltzmann Models}.
\newblock Springer Verlag, 2000.

\bibitem{Guo-Shu:2013}
Z.~L. Guo and C.~Shu.
\newblock {\em Lattice Boltzmann Method and Its Applications in Engineering}.
\newblock World Scientific Publishing, 2013.

\bibitem{Kruger.etc:2017}
T.~Kr\"{u}ger, H.~Kusumaatmaja, A.~Kuzmin, O.~Shardt, G.~Silva, and E.~M.
  Viggen.
\newblock {\em The Lattice Boltzmann Method: Principles and Practice}.
\newblock Springer Verlag, 2017.

\bibitem{zhou.bk.2004}
J.~G. Zhou.
\newblock {\em Lattice Boltzmann Methods for Shallow Water Flows}.
\newblock Springer-Verlag, Berlin, 2004.

\bibitem{Ghia.etc:1982}
U.~Ghia, K.N. Ghia, and C.T. Shin.
\newblock High-{Re} solutions for incompressible flow using the {Navier-Stokes}
  equations and a multigrid method.
\newblock {\em Journal of Computational Physics}, 48:387--411, 1982.

\bibitem{Wong.Baker:2002}
K.~L. Wong and A.~J. Baker.
\newblock A 3d incompressible navier–stokes velocity–vorticity weak form 2nite
  element algorithm.
\newblock {\em International Journal for Numerical Methods in Fluids},
  38:99--123, 2002.

\bibitem{Jiang.etc:1994}
B.~N. Jiang, T.~L. Lin, and L.~A. Povinelli.
\newblock Large scale computation of incompressible viscous flows by
  least-squares finite element method.
\newblock {\em Computer Methods in Applied Mechanics and Engineering},
  114:213--231, 1994.

\end{thebibliography}

\end{document}